\documentclass[aps,reprint,showpacs,twocolumn,superscriptaddress,longbibliography,floatfix]{revtex4-2}
\usepackage[colorlinks,linkcolor=blue,citecolor=blue,anchorcolor=blue,urlcolor=blue]{hyperref}

\usepackage{enumitem}
\usepackage{bm,amssymb,amsmath}
\usepackage{dcolumn}
\usepackage{graphicx}
\usepackage{multirow}
\usepackage{graphicx}

\begin{document}
\title{Fractal depth-first search paths in statistical physics models}
\date{\today}

\author{Qiyuan Shi}
\affiliation{Department of Modern Physics, University of Science and Technology 
of China, Hefei, Anhui 230026, China}

\author{Youjin Deng}
\email{yjdeng@ustc.edu.cn}
\affiliation{Department of Modern Physics, University of Science and Technology 
of China, Hefei, Anhui 230026, China}
\affiliation{Hefei National Research Center for Physical Sciences at the Microscale, 
University of Science and Technology of China, Hefei, Anhui 230026, China}
\affiliation{Hefei National Laboratory, University of Science and Technology of 
China, Hefei, Anhui 230088, China}

\author{Ming Li}
\email{lim@hfut.edu.cn}
\affiliation{School of Physics, Hefei University of Technology, Hefei, 
Anhui 230009, China}

\begin{abstract}
We study the fractal properties of depth-first search (DFS) paths in critical configurations of 
statistical physics models, including the two-dimensional $O(n)$ loop model for various $n$, 
and bond percolation in dimensions $d = 2$ to $6$. In the $O(n)$ loop model, across both 
critical and tricritical Potts regimes, the fractal dimension of the DFS path consistently 
follows $d_{\rm DFS} = 1 + g/8$, where $g$ is the coupling constant in Coulomb gas theory, 
related to $n$ via $n^2 = 2 + 2 \cos(\pi g/2)$ with $g \in [8/3, 16/3]$. For bond percolation, 
the DFS path exhibits nontrivial fractal scaling across all studied dimensions. Interestingly, 
when DFS is applied to the full lattice without any dilution or criticality, the path is still 
fractal in two dimensions, with a dimension close to $7/4$, but becomes space-filling in higher 
dimensions. Our results demonstrate that DFS offers a robust and broadly applicable geometric 
probe for exploring critical phenomena beyond traditional observables.
\end{abstract}

\maketitle

\section{Introduction}     \label{sec-intro}

Phase transitions and critical phenomena~\cite{ElementsofPhase} are central topics in statistical 
physics, where systems exhibit abrupt changes in macroscopic behavior driven by the continuous 
variation of external parameters such as temperature. Near the critical point, physical quantities 
often exhibit singular behavior, such as the divergence of susceptibility. These singularities 
typically follow power-law scaling, reflecting the collective behavior of many interacting 
components and giving rise to universal scaling laws. These phenomena are characterized not 
only by thermodynamic singularities, but also by the emergence of intricate geometric 
structures~\cite{SHavlin_1984,PhysRevLett.56.545,HEStanley_1977,PhysRevLett.53.1121,
nienhuis1984critical,PhysRevLett.62.3054,PhysRevLett.83.1359,Saberi_2009}, such as 
long-range correlations, fractal interfaces, and scale-invariant domains.

The geometric perspective offers a powerful and complementary framework for understanding 
critical phenomena. It has proven essential both for theoretical analysis and for the 
development of efficient Monte Carlo algorithms. A prominent example is the 
Fortuin-Kasteleyn (FK) representation~\cite{kasteleyn1969phase,fortuin1972random} of 
the Potts model~\cite{RevModPhys.54.235}, which enables the formulation of cluster 
algorithms such as the Swendsen-Wang~\cite{PhysRevLett.58.86} and Wolff 
algorithms~\cite{PhysRevLett.62.361}. These methods dramatically improve simulation 
efficiency by reducing critical slowing down and enabling updates of entire correlated 
regions at once.

Moreover, geometric considerations have played a key role in the exact solutions of 
statistical models~\cite{BaxterExactly}. In two dimensions (2D), the scaling limits 
of critical interfaces in models such as percolation~\cite{stauffer1992} and the Ising 
model~\cite{RevModPhys.39.883} can be rigorously described using the Schramm-Loewner 
evolution~\cite{CARDY200581,Smirnov2001,chelkak2014}, which captures the stochastic 
geometry of conformally invariant curves. Parallel to this, conformal field theory 
provides a powerful analytic framework for understanding the scaling behavior of 
critical systems in 2D~\cite{Cardy1987,Friedan1984,Duplantier1989}, enabling the 
exact determination of scaling dimensions and critical exponents. These developments 
collectively highlight the deep interplay between geometry, criticality, and symmetry 
in low-dimensional systems.

A prototypical example of geometric structures at criticality is the fractal of 
clusters, such as those formed in percolation or spin domains in magnetic systems. 
Specifically, the size of a critical cluster, typically defined as the number of 
sites it contains, scales with the linear system size $L$ as $C \sim L^{d_f}$ for 
$L\to\infty$, where $d_f$ is the fractal dimension.

Beyond the cluster size, finer geometric features of critical clusters offer richer 
insights into universality. In 2D, cluster boundaries also exhibit fractal behavior, 
characterized by the so-called hull and external perimeter dimensions, $d_{\rm hull}$ 
and $d_{\rm EP}$~\cite{RMZiff_1984,PhysRevLett.56.545,PhysRevLett.58.2325,
TGrossman_1986,TGrossman_1987}. These exponents capture the roughness of interfaces 
and admit exact expressions in the $Q$-state Potts model via Coulomb gas 
techniques~\cite{nienhuis1984critical,nienhuis1987coulomb,PhysRevLett.62.3054,
PhysRevLett.58.2325,PhysRevLett.82.3940}, 
\begin{align}
d_f &= \frac{(g+2)(g+6)}{8g},    \label{eq-df} \\
d_{\rm EP} &= 1 + \frac{g}{8},   \label{eq-dep}\\
d_{\rm hull} &= 1 + \frac{2}{g},  \label{eq-dhull}
\end{align}
where the Coulomb-gas coupling strength $g$ is related to $Q$ by
\begin{equation}
Q = 2 + 2 \cos(\pi g/2), \quad g \in [2,4]. \label{eq-Qg}
\end{equation}

Additional important geometric observables include the backbone~\cite{PhysRevLett.53.1121} 
and the shortest path~\cite{HJHerrmann_1984}. The backbone, obtained by removing all 
bridge bonds from a cluster~\cite{PhysRevE.69.026114,PhysRevE.89.012120,PhysRevE.105.044122}, 
is itself a self-similar object. Its fractal dimension was recently determined exactly 
in the 2D Potts model using the conformal loop ensemble~\cite{Nolin2023,PhysRevLett.134.117101}. 
Unlike previously known exactly solved critical exponents, the backbone exponent is a 
transcendental number, being a root of an elementary equation. For 2D percolation, this 
yields $d_{\rm B}=1.6433332...$. In comparison, the shortest-path exponent $d_{\rm min}$, 
which governs the scaling of the chemical distance $S \sim r^{d_{\rm min}}$ between two 
connected sites separated by a Euclidean distance $r$, has so far only been determined 
numerically, with the most precise estimate for 2D percolation being 
$d_{\rm min}=1.13077(2)$~\cite{PhysRevE.86.061101}. 
For dimensions $d > 2$ or other models, where no exact solutions are available, 
the determination of such geometric exponents still relies on large-scale numerical 
simulations.

\begin{figure}
\centering
\includegraphics[width=\linewidth]{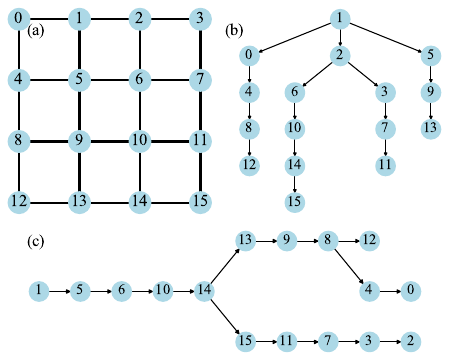}
\caption{(Color online) A schematic illustration of BFS and DFS on a square lattice. 
(a) A $4\times4$ square lattice with sites labeled from $0$ to $15$. (b) A possible 
BFS spanning tree starting from site $1$. The tree depth is $5$, representing the 
shortest path from the initial site $1$ to the farthest site $15$. (c) A possible 
DFS spanning tree starting from site $1$. The tree depth is $9$. In general, the 
DFS path is longer and more circuitous than the BFS path.}
\label{fig-bfsdfs}
\end{figure}

To numerically identify geometrical structures in clusters, two fundamental graph 
traversal algorithms are commonly employed: breadth-first search (BFS) and 
depth-first search (DFS)~\cite{cormen2022introduction}. BFS explores clusters layer 
by layer from a seed site, producing a nearly balanced spanning tree 
(see Fig.~\ref{fig-bfsdfs} (b) for an example), and is particularly well suited for 
evaluating shortest paths and estimating $d_{\rm min}$. In contrast, DFS proceeds 
as far as possible along each branch before backtracking, generating more circuitous 
paths and a distinct tree structure (see Fig.~\ref{fig-bfsdfs} (c)). DFS can also be 
extended to extract specific geometrical features, such as the backbone, by tracking 
and removing bridge edges during traversal~\cite{HJHerrmann_1984,PhysRevE.69.026114,
PhysRevE.89.012120,PhysRevE.105.044122}. While both BFS and DFS are sufficient 
for basic node traversal, identifying specific geometrical structures such as hulls 
or external perimeters requires tailored traversal schemes~\cite{RMZiff_1984,RFVoss_1984,
TGrossman_1986,PGrassberger_1986,asikainen2003fractal,Adams_2010,zatelepin2010}.

In this paper, we demonstrate that DFS paths themselves exhibit fractal scaling. Through 
extensive simulations, we investigate the DFS paths within critical FK clusters of the 2D 
$Q$-state Potts model, realized via the $O(n)$ loop model with $Q = n^2$. 
We find that the fractal dimension $d_{\rm DFS}$ of the DFS path consistently coincides 
with the values given by Eq.~(\ref{eq-dep}) for both critical and tricritical Potts models, 
although Eq.~(\ref{eq-dep}) was originally derived only for the external perimeter in the 
critical Potts model. 
We further extend our analysis to bond percolation on hypercubic lattices 
in dimensions $d = 2$ to $6$. The DFS paths remain fractal even in higher dimensions ($d > 2$), 
for which the notion of an external perimeter has not been generalized. Remarkably, even in 
the absence of any criticality, as in a full lattice, the DFS still generates a nontrivial 
fractal path in 2D, with $d_{\rm DFS} = 1.753(5)$, suggesting a value of $7/4$. In contrast, 
for $d > 2$, the DFS path becomes space-filling, with $d_{\rm DFS} = d$. These results 
identify DFS as a simple yet powerful probe of fractal geometry in statistical physics 
systems, revealing universal features across different models and spatial dimensions. 
We also provide an argument that the DFS paths are dual to the hulls of FK clusters, 
leading to a consistent interpretation of the $d_{\rm DFS}$ values found in our study.

The remainder of this paper is organized as follows. Section~\ref{sec-model} introduces 
the models and simulation algorithms. Section~\ref{sec-results} presents the numerical 
results. A brief discussion is given in Sec.~\ref{sec-con}.

\begin{table*}
\centering
\caption{Fitted results for the fractal dimension of the DFS path in the $O(n)$ loop model 
with various values of $n = \sqrt{Q}$. 
The relation between the Coulomb-gas coupling strength $g$ and $Q$ is determined 
by Eq.~(\ref{eq-Qg}). The columns labeled $\ell_{\rm max}$, $\overline{\ell}_1$, and 
$N_1$ list the fractal dimensions obtained from the corresponding observables, 
while $d_{\rm DFS}$ and $d_f$ denote the values from $d_{\rm DFS} = 1 + g/8$ 
and $d_f = (g + 2)(g + 6)/(8g)$ for the given $g$. The results show that both 
the maximum DFS path length $\ell_{\rm max}$ and the mean DFS path length 
$\overline{\ell}_1$ in the largest cluster scale as $\ell_{\rm max} \sim 
\overline{\ell}_1 \sim L^{d_{\rm DFS}}$, with $d_{\rm DFS} = 1 + g/8$. 
In addition, the number of DFS paths $N_1$ in the largest cluster scales 
proportionally to the cluster size, i.e., $N_1 \sim L^{d_f}$, where 
$d_f = (g + 2)(g + 6)/(8g)$.}
\begin{ruledtabular}
\begin{tabular}{clllllll}
      & $g$  &  $Q$ & $d_{\rm DFS}$  & $d_{f}$ & $\ell_{\rm max}$ & $\overline{\ell}_1$  & $N_1$  \\
    \hline
        & 8/3   &  1  &   4/3      &  91/48      & 1.33319(8)  & 1.333(1)  &   1.8957(2)        \\
        &2.83914& 1.5 &   1.35489  &  1.88322     & 1.35470(11)   & 1.35525(12) &  1.8828(8)        \\
 $x_-$  & 3     &  2  &   11/8     &  15/8     & 1.3746(8)    & 1.3749(4)   &    1.8749(1)      \\
        &3.16086& 2.5 &   1.39511  &  1.86966     & 1.3951(4)    & 1.3951(4) &   1.8693(9)           \\
        & 10/3  &  3  &   17/12    &  28/15     & 1.4156(8)    & 1.4160(8)  &    1.8662(7)        \\
        &3.53989& 3.5 &   1.44249  &  1.86623     & 1.4427(4)     & 1.4429(8)  & 1.8660(8)       \\
   \hline
$x_-=x_+$  & 4     &  4  &   3/2   &  15/8        & 1.4975(27)   &  1.4989(15)  &   1.872(4) \\
   \hline
        &4.46011& 3.5 &   1.55751  &  1.89383     & 1.5573(3)    &  1.5580(7)     &  1.8935(4)      \\
        & 14/3  &  3  &   19/12    &  40/21     & 1.5836(2)    &  1.5830(14)      &  1.9047(4)     \\
        &4.83914& 2.5 &   1.60489  &  1.91486     &  1.6047(3)   &  1.6059(5)  &   1.9147(6)       \\
 $x_+$  & 5     &  2  &   13/8     &  77/40     &  1.6245(3)   &  1.626(4)     &   1.923(2)       \\
        &5.16086& 1.5 &   1.64511  &  1.93576     &  1.6446(3)   & 1.646(2)    &    1.9362(7)        \\
        & 16/3  &  1  &   5/3      &  187/96     & 1.6663(6)    &  1.6666(9)     &   1.9476(7)    \\
\end{tabular} \label{tab1}
\end{ruledtabular}
\end{table*}

\section{Models, algorithm, and obervables} \label{sec-model}

\subsection{$O(n)$ loop model}

To explore a general statistical framework, we consider the $Q$-state Potts 
model, which reduces to the percolation model at $Q=1$ and the Ising model 
at $Q=2$. However, for large $Q$, Monte Carlo simulations of the Potts model 
suffer from severe critical slowing down and poor convergence of finite-size 
corrections~\cite{PhysRevLett.63.827}. To overcome these issues, we adopt the 
$O(n)$ loop model~\cite{DOMANY1981279} on the honeycomb lattice with the algorithm 
introduced in Ref.~\cite{PhysRevE.105.044122}, which corresponds to the critical 
and tricritical Potts models with $Q = n^2$ in the branches $x_-$ and $x_+$, 
respectively~\cite{PhysRevLett.49.1062,NIENHUIS1991109}, where
\begin{equation}
x_{\pm} = \frac{1}{\sqrt{2 \pm \sqrt{2 - n}}}.
\end{equation}

Following the algorithm introduced in Ref.~\cite{PhysRevE.105.044122}, the $O(n)$ loop 
model on the honeycomb lattice can be efficiently simulated on the triangular lattice 
as a generalized Ising model. In this representation, each site carries an Ising spin 
and an additional binary state indicating whether it is active or not. Starting from 
an Ising spin configuration, the system evolves iteratively as follows:
\begin{enumerate}
  \item For each Ising domain, all sites are independently set to be active with 
  probability $1/n$; otherwise, all are inactive.
  \item Bonds are then occupied according to the following rules:
  \begin{itemize}
    \item For two adjacent active sites with the same spin, occupy the bond between 
    them with probability $1 - x$, where $x = x_-$ or $x_+$.
    \item For inactive sites, occupy all adjacent bonds, regardless of the state 
    of the neighboring site.
  \end{itemize}
  \item For each cluster connected by the occupied bonds, flip the Ising spin of 
  all sites in the cluster with probability $1/2$ (note that spins within a cluster 
  may differ). The resulting Ising domains have the same geometrical structure 
  as the FK clusters of the critical ($x_-$) or tricritical ($x_+$) Potts model with $Q = n^2$.
\end{enumerate}
Using this algorithm, we simulate the $O(n)$ loop model on the honeycomb lattice 
(generalized Ising model on the triangular lattice) with periodic boundary conditions, 
with linear system sizes $L$ ranging from $16$ to $4096$.

\subsection{Percolation model}

In addition to the 2D $O(n)$ loop model, we also investigate critical bond percolation 
on hypercubic lattices in dimensions $d = 2$ to $6$. In this model, each bond of the 
lattice is independently occupied with probability $p$ and left unoccupied with 
probability $1-p$. At the percolation threshold $p = p_c$, the system undergoes a 
continuous phase transition, and large-scale connected clusters emerge with 
nontrivial fractal geometry.

We generate critical configurations by setting $p = p_c$, using the exact solution 
$p_c = 1/2$ for $d=2$, and the numerically established values 
$p_c = 0.24881182(10)$~\cite{PhysRevE.87.052107} for $d=3$, 
$p_c = 0.1601312(2)$~\cite{PhysRevResearch.2.013067} for $d=4$, 
$p_c = 0.11817145(3)$~\cite{PhysRevE.98.022120} for 
$d=5$, and $p_c = 0.09420165(2)$~\cite{PhysRevE.98.022120} for $d=6$.

\subsection{DFS process and measured quantities}

To probe the scaling behavior of DFS paths, we first generate critical configurations as 
described above. For each configuration, we perform DFS on every cluster to extract the DFS 
paths. The DFS algorithm proceeds as follows. Starting from a randomly chosen site 
(e.g., site 1 in Fig.~\ref{fig-bfsdfs} (a)), we mark the site as visited and recursively 
explore its unvisited neighbors. At each step, we randomly choose one unvisited neighbor 
and move to it, continuing the process until reaching a site with no unvisited neighbors. 
This traversal creates a chain of visited sites, referred to as a DFS path (e.g., the 
path from $1$ to $2$ in Fig.~\ref{fig-bfsdfs} (c)). Then, at this point, the algorithm 
backtracks to the previous site and continues exploring any remaining unvisited neighbors. 
The process continues until all reachable sites in the cluster have been visited. Note that 
multiple DFS paths can be identified during the traversal, and they may share some common 
sites and bonds (e.g., paths from $1$ to $12$, and from $1$ to $0$ in Fig.~\ref{fig-bfsdfs} (c)). 
We emphasize that in our DFS process, the neighbors are visited in a random order. A 
fixed visiting order can lead to biased paths that depend on the specific order and lattice 
structure, particularly for dense clusters (e.g., the $O(n)$ loop model on the $x_+$ branch, 
or the $x_-$ branch with large $Q$).

In the simulations, we mainly focus on the length $\ell$ of these DFS paths. In particular, 
for a given configuration, we measure the following quantities:
\begin{enumerate}
\item The maximum length $\ell_{\rm max}$ of all DFS paths.
\item The mean length $\overline{\ell}_1$ of all DFS paths in the largest cluster.
\item The number $N_1$ of DFS paths found in the largest cluster.
\item The number density of the length of the longest DFS path in each clusters, 
denoted as $P(\ell) = \mathcal{N}(\ell', \Delta \ell)/L^d \Delta \ell$, where 
$\mathcal{N}(\ell', \Delta \ell)$ is the number of clusters whose longest DFS 
path lies in the interval $[\ell', \ell' + \Delta \ell)$. To improve statistics 
over a broad range of $\ell$, we choose the bin size as $\Delta \ell = a^k$ 
for the $k$-th bin starting from $\ell = 1$, where $a > 1$ is an adjustable 
parameter controlling the total number of bins. The representative length of 
each bin is then taken as the geometric mean $\ell=\sqrt{\ell'(\ell' + \Delta \ell)}$.
\item The length-resolved count of DFS paths in the largest cluster, denoted as 
$P_1(\ell) = \mathcal{N}_1(\ell', \Delta \ell)/\Delta \ell$, where 
$\mathcal{N}_1(\ell', \Delta \ell)$ is the number of DFS paths in the 
largest cluster whose lengths fall in $[\ell', \ell' + \Delta \ell)$. The 
representative length $\ell=\sqrt{\ell'(\ell' + \Delta \ell)}$ and the binning 
scheme are chosen in the same way as for $P(\ell)$.
\end{enumerate}
We remark that the above quantities are defined for a single configuration. 
In practice, all measurements are obtained by performing ensemble averages over 
many independent configurations. For simplicity of notation, we use the same 
symbols (e.g., $\ell_{\rm max}$, $\overline{\ell}_1$, etc.) to denote the 
ensemble-averaged values, without introducing separate symbols for the averages.

\section{Simulation results} \label{sec-results}

\subsection{ $O(n)$ loop model}

\begin{figure}[bhtp]
\centering
\includegraphics[width=\linewidth]{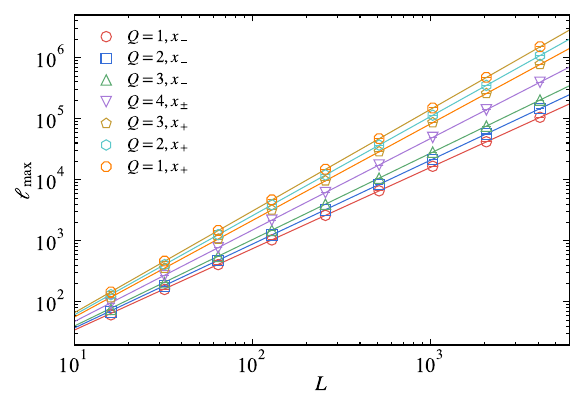}
\caption{(Color online) Finite-size scaling of the maximum DFS path length $\ell_{\rm max}$ 
in the $O(n)$ loop model for various values of $n$. The solid lines represent the 
fractal dimensions given by Eq.~(\ref{eq-dep}). Note that Eq.~(\ref{eq-dep}) was originally 
derived for the external perimeter in the critical Potts model; here, the scaling of the 
DFS path is also consistently described by the same expression for both the critical and 
tricritical Potts models.} \label{fig-fsson}
\end{figure}

\subsubsection{Fractal dimension of DFS paths}

We begin by examining the finite-size scaling behavior of DFS paths in the $O(n)$ loop model. 
As shown in Fig.~\ref{fig-fsson}, the maximum DFS path length $\ell_{\rm max}$ exhibits a 
clear power-law dependence on the system size $L$,
\begin{equation}
\ell_{\rm max} \sim L^{d_{\rm DFS}}, \label{eq-ddfs}
\end{equation}
indicating that the DFS path is a fractal object characterized by a well-defined scaling 
exponent $d_{\rm DFS}$.

To determine $d_{\rm DFS}$, we fit the Monte Carlo data of $\ell_{\rm max}$ to the standard 
finite-size scaling ansatz,
\begin{equation}
Q(L) = L^{d_Q}(a_0 + a_1 L^{-y_1} + \cdots), \label{eq-fss}
\end{equation}
where $d_Q$ describes the leading scaling behavior, $y_1$ accounts for subleading corrections, 
and the ellipsis denotes higher-order terms. During the fitting procedure, we apply a lower 
cutoff $L \geq L_{\rm min}$ and assess the goodness of fit by examining $\chi^2$. The 
optimal $L_{\rm min}$ is chosen as the smallest value beyond which $\chi^2$ per degree 
of freedom (DF) no longer decreases significantly (more than one unit) as $L_{\rm min}$ 
increases. We consider a fit acceptable if $\chi^2/ \text{DF} \approx 1$. Systematic 
uncertainties are estimated by varying $y_1$, including both fixed and free values. 
The fit results are summarized in Table~\ref{tab1}. 

\begin{figure}
\centering
\includegraphics[width=\linewidth]{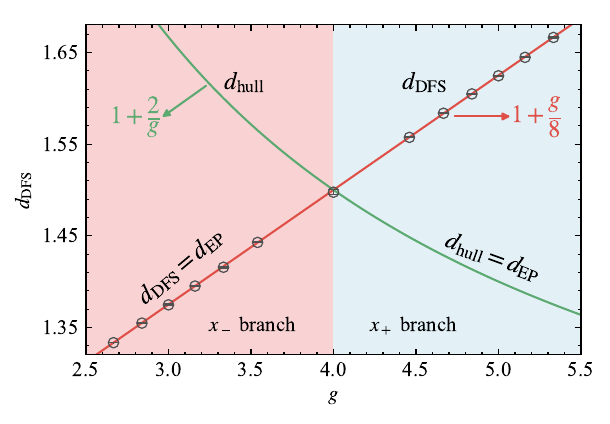}
\caption{(Color online) Numerical results for the fractal dimension $d_{\rm DFS}$ of the $O(n)$ 
loop model as a function of the Coulomb-gas coupling strength $g$. The red and blue regions 
correspond to the $x_-$ and $x_+$ branches, respectively. 
The DFS and hull dimensions are given by Eqs.~(\ref{eq-dep}) and (\ref{eq-dhull}) (red and green 
lines). The external-perimeter dimension coincides with $d_{\rm DFS}$ in the critical Potts 
model ($x_-$ branch) and with $d_{\rm hull}$ in the tricritical Potts model ($x_+$ branch). 
The scattered points represent our numerical fit results for $d_{\rm DFS}$, which are consistent 
well with Eq.~(\ref{eq-dep}).} \label{fig-gvsd}
\end{figure}

In Fig.~\ref{fig-gvsd}, we plot the fitted fractal dimensions $d_{\rm DFS}$ as a 
function of the Coulomb-gas coupling constant $g$. The numerical values of $d_{\rm DFS}$ 
for both the $x_-$ and $x_+$ branches exhibit excellent agreement with the expression in 
Eq.~(\ref{eq-dep}), i.e., $d_{\rm DFS}=1+g/8$, even though that expression was not 
originally derived for DFS paths.

The agreement observed in the $x_-$ branch is nontrivial yet understandable. 
Equation~(\ref{eq-dep}) was derived for the external perimeter dimension $d_{\rm EP}$ of 
the critical Potts model ($g\in[2,4]$), while our quantity is the fractal dimension of 
the longest DFS path. In the $x_-$ branch FK clusters are sparse, and the longest DFS 
path tends to explore the outer boundary of a cluster without entering every microscopic 
indentation. Due to this boundary following and the skip over of fine-scale fjord, its 
geometry is expected to be that of an external perimeter rather than the hull or the full cluster.

The remarkable observation, however, arises in the $x_+$ branch, where FK clusters are 
dense. The geometry of these dense clusters leaves no deep fjords, so that the hull itself 
effectively serves as the external perimeter. Numerical results in Ref.~\cite{PhysRevE.105.044122} 
further show that FK clusters in the $x_+$ branch have identical backbone and mass dimensions 
($d_B = d_f$), providing additional evidence of their compactness. Consequently, the two 
boundaries become indistinguishable, and their common dimension is given by the hull 
dimension $d_{\rm hull}$ from Eq.~(\ref{eq-dhull}), as indicated by the green line in 
Fig.~\ref{fig-gvsd}.

The persistence of this agreement in the dense phase, as shown in Fig.~\ref{fig-gvsd}, 
strongly suggests that the DFS path is not merely tracing the external perimeter but is instead 
governed by a more fundamental geometric correspondence, which holds irrespective of the bond 
density of the cluster.

To provide a more general understanding of the observed behavior, we propose that the 
fractal DFS paths can be viewed as dual counterparts of the hulls. In this picture, the fractal 
dimensions of the DFS paths, $d_{\rm DFS}$, and of the hulls, $d_{\rm hull}$, can be described 
by the same functional form in terms of the Coulomb-gas coupling strength, namely 
Eq.~(\ref{eq-dhull}), where the coupling strengths are related by the duality 
condition $g_{\rm DFS}\cdot g_{\rm hull}=16$. Formally, this implies that $d_{\rm DFS}$ can be 
obtained from Eq.~(\ref{eq-dhull}) by replacing $g$ with the dual coupling $16/g$, which is 
just Eq.~(\ref{eq-dep}).

Therefore, due to this duality, the relative magnitude between $d_{\rm DFS}$ and 
$d_{\rm hull}$ is inverted at $g_{\rm DFS} = g_{\rm hull} = 4$, corresponding to $n=2$ 
(see Fig.~\ref{fig-gvsd}). Qualitatively, along the $x_-$ branch (critical Potts model), 
FK clusters are sparse and their hulls highly tortuous. A DFS path cannot fully penetrate 
the fine-scale fjords and thus appears relatively smoother (i.e., more like a simple line), 
leading to $d_{\rm DFS} < d_{\rm hull}$. In contrast, along the $x_+$ branch (tricritical 
Potts model), since the FK cluster is dense (with relatively smoother hulls), a DFS path 
has more freedom to wander inside the cluster, becoming less line-like and more 
tortuous, yielding $d_{\rm DFS} > d_{\rm hull}$. This scenario further suggests that, in 
the extreme case of a fully occupied lattice corresponding to a maximally dense cluster, 
the same duality between DFS paths and hulls is expected to hold (see Sec.~\ref{sec-fl}).

\begin{figure*}[htp]
\centering
\includegraphics[width=\linewidth]{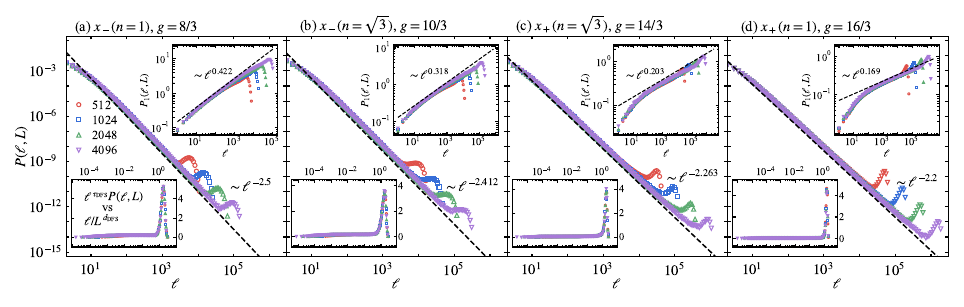}
\caption{(Color online) The number density $P(\ell,L)$ of the length of the longest DFS 
path in each cluster for the $O(n)$ loop model with different values of $n$ and linear 
size $L$. Panels (a) and (b) correspond to the $x_-$ branch, while (c) and (d) correspond 
to the $x_+$ branch. The dashed lines represent the Fisher exponent predicted by 
hyperscaling relation, $\tau_{\rm DFS} = 1 + d/d_{\rm DFS}$. The upper 
insets show the length-resolved count $P_1(\ell,L)$ of DFS paths in the largest cluster, 
which also exhibits a power-law form, $P_1(\ell,L) \sim \ell^{\tau_1}$, with $\tau_1 = 
d_f/d_{\rm DFS} - 1$. The lower insets show the scaling collapse plots of 
$\ell^{\tau_{\rm DFS}} P(\ell,L)$ versus $\ell/L^{d_{\rm DFS}}$, with $d_{\rm DFS}=1+g/8$.}
\label{fig-P}
\end{figure*}

\subsubsection{Length distribution of DFS paths}

The self-similar nature of criticality also implies that the length distribution of 
the longest DFS path in each cluster follows a power law,
\begin{equation}
P(\ell) \sim \ell^{-\tau_{\rm DFS}},
\end{equation}
where the Fisher exponent $\tau_{\rm DFS}$ satisfies the hyperscaling relation
\begin{equation}
\tau_{\rm DFS} = 1 + \frac{d}{d_{\rm DFS}}. \label{eq-taudfs}
\end{equation}
This scaling is also observed for both the $x_-$ and $x_+$ branches, as demonstrated 
in Fig.~\ref{fig-P}. Plotting $\ell^{\tau_{\rm DFS}}P(\ell)$ as a function 
of $\ell/L^{d_{\rm DFS}}$ results in a satisfactory collapse across different system 
sizes (see the lower insets), confirming the fractal nature of the DFS paths and the 
corresponding $g$-dependent fractal dimension.

Moreover, the fractal nature is not limited to the longest DFS path. The mean DFS path length 
in the largest cluster, denoted by $\overline{\ell}_1$, follows the same scaling form in 
Eq.~(\ref{eq-ddfs}). As listed in Table~\ref{tab1}, the fitted fractal dimension $d_{\rm DFS}$ 
from the data of $\overline{\ell}_1$ agrees with that obtained from $\ell_{\rm max}$.

To further characterize the structure of DFS paths, we define $P_1(\ell)$ as the length-resolved 
count of DFS paths in the largest cluster. As shown in the insets of Fig.~\ref{fig-P}, 
$P_1(\ell)$ exhibits a power-law form for increasing $L$,
\begin{equation}
P_1(\ell) \sim \ell^{\,\tau_1}.
\end{equation}
We also find that the total number of DFS paths in the largest cluster, $N_1$, scales proportionally 
to the cluster size $C_1$, i.e., $N_1 \sim C_1 \sim L^{d_f}$ (see Table~\ref{tab1}). Therefore, 
integrating the distribution $P_1(\ell)$ yields
\begin{equation}
\int_1^{L^{d_{\rm DFS}}} \ell^{\tau_1} d\ell \sim L^{d_f},
\end{equation}
from which we obtain the scaling relation
\begin{equation}
\tau_1 = \frac{d_f}{d_{\rm DFS}} - 1.
\end{equation}
This prediction is confirmed by the observed $\tau_1$ values shown in the upper insets 
of Fig.~\ref{fig-P}.

\subsection{Bond percolation model}

\begin{table}
\centering
\caption{Fitting results for the fractal dimension of the DFS path, $d_{\rm DFS}$, in bond 
percolation on hypercubic lattices in dimensions $d = 2$ to $6$. The results are obtained by 
fit the data of $\ell_{\rm max}$ to the finite-size scaling anastz Eq.~(\ref{eq-fss}). 
Parameters shown without error bars are fixed during the fitting procedure.}
\begin{ruledtabular}
\begin{tabular}{crlllll}
 $d$    & $L_{\rm min}$  & $d_{\rm DFS}$  & $a_0$ &  $y_1$ & $a_1$    &  $\chi^2/\rm{DF}$     \\
\hline
2       & 32   &  1.3334(1)  &  1.545(1)  &   1.34(6)  &  -1.4(3)  &  8.61/6   \\
        & 64   &  1.3333(3)  &  1.546(3)  &   1.2(2)   &  -0.9(7)  &  8.05/5   \\
        & 128  &  1.3332(2)  &  1.546(2)  &   1.2      &  -0.9(1)  &  7.91/5   \\
\hline
3       & 24   &  1.5757(5)  &  1.592(4)  &   1        &  -0.11(3) &  7.73/5   \\
        & 32   &  1.5754(8)  &  1.594(6)  &   1        &  -0.13(6) &  7.42/4   \\
        & 64   &  1.5759(5)  &  1.589(3)  &   -        &  -        &  4.80/3   \\
\hline
4       & 5    &  1.7415(4)  &  1.865(3)  &   2.47(4)  &  -4.4(2)  &  10.80/6   \\
        & 7    &  1.7426(5)  &  1.858(3)  &   2.7(1)   &  -7(1)    &  4.95/5   \\
        & 10   &  1.7422(4)  &  1.861(2)  &   2.5      &  -4.3(2)  &  5.72/5   \\
\hline
5       & 6    &  1.86(2)    &  2.6(2)    &  -0.9(3)   &  -1.4(1)  &  8.19/7   \\
        & 6    &  1.863(1)   &  2.58(1)   &  -1        &  -1.48(4) &  8.26/8   \\
        & 12   &  1.864(5)   &  2.58(5)   &  -1        &  -1.5(2)  &  8.24/7   \\
\hline
6       & 4  &  2.077(3)  &  2.77(3)  &  -2.01(7)  &  -5.3(3)  &  8.90/8   \\
        & 5  &  2.074(6)  &  2.79(5)  &  -1.9(2)   &  -4.9(7)  &  8.50/7   \\
        & 6  &  2.077(2)  &  2.77(2)  &  -2        &  -5.2(2)  &  8.73/7   \\
\end{tabular} \label{tab2}
\end{ruledtabular}
\end{table}

\begin{figure}
\centering
\includegraphics[width=\linewidth]{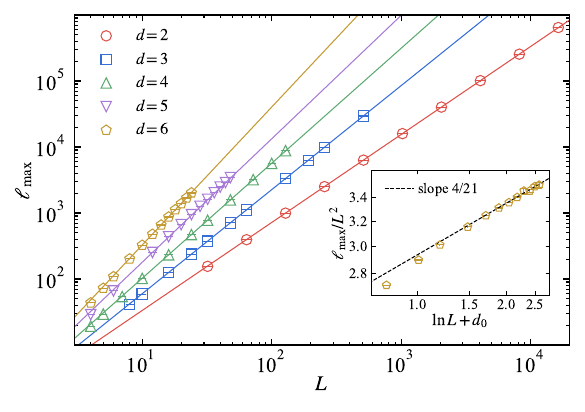}
\caption{(Color online) Finite-size scaling of the maximum DFS path length $\ell_{\rm max}$ in 
the bond percolation model for different spatial dimensions $d$. In all dimensions, a clear 
power-law scaling $\ell_{\rm max} \sim L^{d_{\rm DFS}}$ is observed, with a $d$-dependent 
fractal dimension $d_{\rm DFS}$. The solid lines represent the fitting results based on the 
scaling ansatz given in Eq.~(\ref{eq-fss}) (Table~\ref{tab2}). 
The inset is a rescaled plot of the data for $d=6$, where $\ell_{\rm max}/L^2$ is plotted 
against $\ln L + d_0$ (with $d_0 = -0.6$) in log-log coordinates.
The apparent straight-line behavior with slope $4/21$ demonstrates a multiplicative logarithmic 
correction of the form $\ell_{\rm max}/L^2 \sim (\ln L + d_0)^{4/21}$.}
\label{fig-fssper}
\end{figure}

We next investigate the finite-size scaling behavior of the DFS path in the bond percolation model 
on hypercubic lattices for dimensions $d=2$ to $6$. As shown in Fig.~\ref{fig-fssper} with fit 
results listed in Table~\ref{tab2}, the maximum DFS path length $\ell_{\rm max}$ again exhibits 
a clear power-law scaling with the system size $L$, $\ell_{\rm max}\sim L^{d_{\rm DFS}}$, 
demonstrating that the DFS path forms a fractal object across dimensions and models.

In particular, for $d=2$, percolation model corresponds to the $Q=1$ case of the Potts model, which 
was also studied in the $O(n)$ loop model (see Table~\ref{tab1} and Fig.~\ref{fig-fsson}). It is 
worth noting that while the $O(n)$ loop model with $Q=1$ corresponds to site percolation on the 
triangular lattice, here we consider bond percolation on the square lattice. Despite this 
difference, the fitted fractal dimensions $d_{\rm DFS}=4/3$ are consistent between site and 
bond percolation models, providing strong evidence for the universality of the fractal nature 
of the DFS path across different percolation types.

More interestingly, for dimensions $d \geq 3$, there exists no established geometric or topological 
definition of the external perimeter and hull. Consequently, neither identification algorithms nor 
numerical characterizations of their fractal properties are currently known. Nevertheless, as 
shown in Table~\ref{tab2}, our results estimate the DFS path dimension $d_{\rm DFS}$ as 
$1.5757(8)$, $1.7426(5)$, $1.864(5)$, and $2.074(6)$ for $d=3$ to $6$, respectively. These 
findings highlight the robustness and universality of DFS path fractality in percolation 
systems, reflecting previously unexplored geometric and topological features of critical 
clusters beyond 2D.

Furthermore, we conjecture that for $d \geq 6$, $d_{\rm DFS} = d_{\rm min} = d/3$
~\cite{PhysRevE.97.022107}, since critical percolation clusters in this regime are 
essentially tree-like, for which the DFS and BFS yield the same spanning tree. For $d=6$, 
the observed deviation $d_{\rm DFS} = 2.074(6) \neq 2$ is likely due to multiplicative 
logarithmic corrections that were not accounted for in the fitting procedure.

In the inset of Fig.~\ref{fig-fssper}, we plot $\ell_{\max}/L^2$ as a function of 
$\ln L+d_0$ with $d_0=-0.6$, and a clear scaling behavior is observed for increasing $L$, 
suggesting $\ell_{\max}/L^2 \sim (\ln L +d_0)^{\hat{d}_{\rm DFS}}$ with $\hat{d}_{\rm DFS}=4/21$. 
This scaling is consistent with the usual expectations for systems at the upper critical dimension, 
and the exponent $4/21$ for the multiplicative logarithmic correction agrees with that of the 
fractal dimension of the percolation cluster~\cite{RuizLorenzo1998}.

\subsection{Full lattices}  \label{sec-fl}

\begin{figure}
\centering
\includegraphics[width=\linewidth]{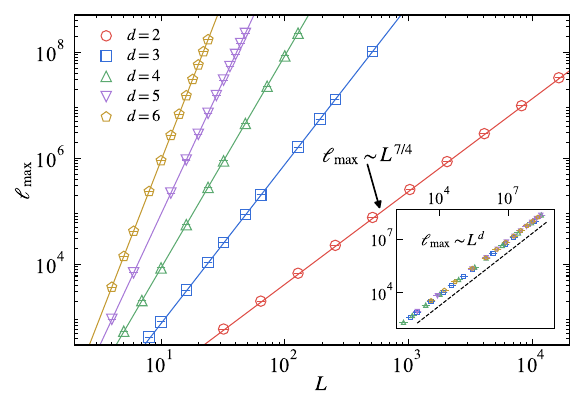}
\caption{(Color online) Finite-size scaling of the maximum DFS path length $\ell_{\rm max}$ on 
entire hypercubic lattices. The scattered points represent the simulation results, 
and the lines denote the scaling $\ell_{\rm max} \sim L^{d_{\rm DFS}}$, with $d_{\rm DFS} = 7/4$ 
for $d=2$ and $d_{\rm DFS} = d$ for $d>2$. The inset shows $\ell_{\rm max}$ as a function 
of $L^d$ for $d>2$, where all data points collapse onto a straight line, confirming the 
trivial scaling behavior in higher dimensions.}
\label{fig-full}
\end{figure}

We further explore DFS behavior on entire hypercubic lattices, where all bonds and sites 
are present by construction, without invoking any percolation process or criticality. 
Interestingly, in 2D, a nontrivial fractal scaling persists: 
$\ell_{\rm max} \sim L^{d_{\rm DFS}}$ with $d_{\rm DFS} = 1.753(5) < d$ 
(see Fig.~\ref{fig-full}). This persistent fractal behavior in 2D is remarkable 
because it emerges without any preconditioned criticality or disorder in the lattice 
structure, hinting at a form of self-organized criticality inherent in the DFS process 
itself. However, for $d>2$, $\ell_{\rm max}$ exhibits trivial scaling as 
$\ell_{\rm max} \sim L^d$, which is well demonstrated by plotting $\ell_{\rm max}$ 
as a function of $L^d$ (see the inset of Fig.~\ref{fig-full}). This indicates that 
the DFS process essentially fills the entire lattice volume.

From the numerical estimate $d_{\rm DFS} = 1.753(5)$ in 2D, one can infer that the fractal 
dimension is consistent with $7/4$. This observation can be understood using our previous 
argument that DFS paths is dual to hulls. Along the $x_+$ branch, FK clusters become increasingly 
dense as $g$ increases ($n\to 0$). When $g=6$, Eq.~(\ref{eq-ddfs}) yields $d_f = 2$, indicating 
that the FK cluster effectively fills the 2D lattice. Therefore, DFS on a full 2D lattice can 
be regarded as a DFS on the FK cluster at $g=6$. According to our duality argument, 
substituting $g=6$ into Eq.~(\ref{eq-dep}) or $g=8/3$ into Eq.~(\ref{eq-dhull}) gives a fractal 
dimension $d_{\rm DFS}=7/4$, consistent with the numerical estimate.

\section{Discussion} \label{sec-con}

In this work, we systematically studied the fractal and scaling properties of DFS paths in various 
statistical physics models, revealing the widespread existence of fractal characteristics in 
these paths. First, for the 2D $O(n)$ loop model, the expression originally developed for 
the external perimeter dimension in the critical Potts model can also describe the fractal 
dimension of DFS paths, for both critical and tricritical Potts regimes, highlighting an intrinsic 
geometric property of the DFS process. Second, in higher-dimensional systems, studied via bond 
percolation on hypercubic lattices, we found that the fractal nature of DFS paths persists, even 
though the concept of an external perimeter is not well-defined beyond 2D. This suggests the 
emergence of new fractal structures and potentially new critical exponents in higher dimensions. 
Third, even on fully occupied 2D lattices, without any dilution or criticality, DFS paths exhibit 
nontrivial fractal scaling of their length, indicating that such fractal geometry can arise 
spontaneously, independent of fine-tuned critical points. Finally, we propose that the 
fractal DFS paths are dual to the hulls in FK clusters, providing a unified framework for 
understanding the fractal geometry of DFS paths across different systems.

These findings point to the fractal nature of DFS paths as a robust and universal feature across 
diverse models and dimensions. This raises compelling questions about the mechanisms driving such 
geometric complexity. On one hand, DFS can be viewed as a particular type of random walk, which is 
history-dependent, loop-suppressing, and self-avoiding. The generated hierarchical and tree-like 
structure leads to distinct scaling behavior that deviates from conventional random walks. On the 
other hand, the DFS path generation can also be interpreted as a cluster growth process, drawing 
potential connections to stochastic growth models such as the Eden model~\cite{eden1961two} and 
the Kardar-Parisi-Zhang universality class~\cite{PhysRevLett.56.889}.

A promising direction for future work lies in establishing analytical frameworks to explain these 
fractal properties, possibly by embedding DFS within broader families of random walks or growth 
processes~\cite{PhysRevLett.81.5489,Lawler2001,Lawler2002}. Exploring such connections may not 
only deepen our understanding of DFS but also reveal new universal aspects of geometry and scaling 
in stochastic systems.

\emph{Note added.} During the review process of this manuscript, we became aware of the 
concurrent work by Soares et al. ~\cite{cthn-nh7b}, which touches upon the topic of DFS on 
percolation clusters from a different perspective.

\section*{Acknowledgment}
The research was supported by the National Natural Science Foundation of China under 
Grant No.~12275263, the Innovation Program for Quantum Science and Technology under 
Grant No.~2021ZD0301900, and Natural Science Foundation of Fujian province of China 
under Grant No.~2023J02032.

\section*{Data availability}
The data that support the findings of this article are openly
available~\cite{shi_2025_16876597}.

\bibliography{ref.bib}

\end{document}